\documentclass[sn-basic]{sn-jnl}% Basic Springer Nature Reference Style/Chemistry Reference Style

\usepackage{graphicx}%
\usepackage{multirow}%
\usepackage{amsmath,amssymb,amsfonts}%
\usepackage{amsthm}%
\usepackage{mathrsfs}%
\usepackage[title]{appendix}%
\usepackage{xcolor}%
\usepackage{textcomp}%
\usepackage{manyfoot}%
\usepackage{booktabs}%
\usepackage{algorithm}%
\usepackage{algorithmicx}%
\usepackage{algpseudocode}%
\usepackage{listings}%
\usepackage{xspace}

\theoremstyle{thmstyleone}%
\theoremstyle{thmstyletwo}%

\theoremstyle{thmstylethree}%

\newcommand{\tess}{\textit{TESS}\xspace}
\newcommand{\ktwo}{\textit{K2}\xspace}
\newcommand{\kepler}{\textit{Kepler}\xspace}
\newcommand{\Corot}{\textit{CoRoT}\xspace}
\newcommand{\cheops}{\textit{CHEOPS}\xspace}
\newcommand{\plato}{\textit{PLATO}\xspace}
\newcommand{\kms}{km\,$\mathrm{s}^{-1}$}

\newcommand{\rev}[1]{{#1}}

\newcommand{\re}[1]{{{#1}}}

\begin{document}

\title[Article Title]{Observations of Exocomets}


\author*[1,2]{\fnm{Judith} \sur{Korth}}\email{judithkorth@gmail.com}

\author[3,4]{\fnm{Azib} \sur{Norazman}}\email{azib.norazman@warwick.ac.uk}

\author[5]{\fnm{Rapha\"el} \sur{Bendahan-West}}\email{rb941@exeter.ac.uk}

\author[3,4]{\fnm{Grant} \sur{Kennedy}}\email{g.kennedy@warwick.ac.uk}
\author[3,4]{\fnm{Cristina} \sur{Madurga Favieres}}\email{cristina.madurga-favieres@warwick.ac.uk}

\author[6]{\fnm{Daniela} \sur{Iglesias}}\email{D.P.Iglesias@leeds.ac.uk}

%\author[x]{\fnm{Geraint}\sur{Jones}}\email{geraint.jones@esa.int}

\author[7,8]{\fnm{Olena}\sur{Shubina}}\email{oshubina@astro.sk}

\author[9]{\fnm{Siyi} \sur{Xu}\email{siyi.xu@noirlab.edu}}

\author[10,11]{\fnm{Nathan X.}\sur{Roth}}\email{nathaniel.x.roth@nasa.gov}


\affil[1]{Lund Observatory, Division of Astrophysics, Department of Physics, Lund University, Box 118, 22100 Lund, Sweden}

\affil[2]{Observatoire astronomique de l’Université de Genève, Chemin Pegasi 51, 1290 Versoix, Switzerland}

\affil[3]{Department of Physics, University of Warwick, Coventry CV4 7AL, UK}

\affil[4]{Centre for Exoplanets and Habitability, University of Warwick, Coventry CV4 7AL, UK}

\affil[5]{Department of Physics and Astronomy, University of Exeter, Stocker Road, Exeter EX4 4QL, UK}

%\affil[2]{\orgdiv{Department}, \orgname{Organization}, \orgaddress{\street{Street}, \city{City}, \postcode{10587}, \state{State}, \country{Country}}}

%\affil[3]{\orgdiv{Department}, \orgname{Organization}, \orgaddress{\street{Street}, \city{City}, \postcode{610101}, \state{State}, \country{Country}}}

\affil[6]{\orgdiv{School of Physics and Astronomy}, \orgname{University of Leeds}, \orgaddress{\street{Sir William Henry Bragg Building}, \city{Leeds}, \postcode{LS2 9JT}, \country{UK}}}

\affil[7]{\orgname{Astronomical Institute of Slovak Academy of Sciences}, \orgaddress{\street{Tatransk\'{a} Lomnica}, \city{Vysok\'{e} Tatry}, \postcode{05960}, \country{Slovak Republic}}}

\affil[8]{\orgname{Main Astronomical Observatory of National Academy of Sciences of Ukraine}, \orgaddress{\street{27 Akademika Zabolotnoho St.}, \city{Kyiv}, \postcode{03143}, \country{Ukraine}}}

\affil[9]{Gemini Observatory/NOIRLab, 950 N Cherry Ave, Tucson, AZ 85719, USA}

\affil[10]{\orgname{NASA Goddard Space Flight Center, Astrochemistry Laboratory Code 691}, \orgaddress{\street{8800 Greenbelt Rd}, \city{Greenbelt}, \state{MD} \postcode{20771}, \country{USA}}}

\affil[11]{\orgname{Department of Physics, American University}, \orgaddress{\street{4400 Massachusetts Ave NW}, \city{Washington}, \state{DC} \postcode{20016}, \country{USA}}}


\abstract{\rev{Active small bodies in extrasolar systems, the extrasolar analogues of Solar System comets, provide insights into the orbital evolution and physical processes shaping planetary systems.} Since the discovery of exocomets around $\beta$\,Pictoris, these small, icy bodies have shown the potential to become key probes for understanding planetary formation and migration. This review presents an overview of current observational techniques used to detect exocomets, focusing on individual systems and large-scale searches. We discuss photometric methods that identify exocomet transits through asymmetric light curves and spectroscopic techniques revealing cometary gases via time-variable absorption lines. Despite progress, significant open questions remain regarding the physical properties, occurrence rates, and similarities between exocomets and their Solar System counterparts. This review explores future opportunities in observational exocomet research, highlighting advancements required to further our understanding of these \rev{active small} bodies and their role in the context of planetary system evolution.}

\keywords{exocomets, photometry, spectroscopy}



\maketitle

\section{Introduction}\label{sec1}

\rev{Active small bodies in extrasolar systems, often referred to as exocomets, are \rev{evolving} and intriguing objects that offer unique insights into the processes occurring in other planetary systems. The term exocomet is used here following the \re{precise} definition proposed by \citet{Iglesias2025}, referring to minor bodies orbiting stars other than the Sun that show signs of sublimation, together with their surrounding tails or comae composed of escaping dust and gas. \re{This definition builds on earlier works \citep[such as][]{2014A&A...561L..10K, 2014Natur.514..462K, Strom2020} that first introduced and applied the term exocomet to describe sublimating, comet-like analogues in extrasolar systems.} As \re{these exocometary bodies} approach their host stars, \re{sublimation} processes can produce prominent tails and comae that render them observable. Since the discovery of the first exocomets around $\beta$\,Pictoris through variable absorption features in its spectrum \citep{Ferlet1987}, exocomets have emerged as valuable tracers of planetary formation and the \rev{orbital} evolution of planetary systems.}

This review focuses on the current \rev{direct} observational techniques used to detect exocomets, their application to individual systems and large-scale searches, and the future possibilities for observational exocomet research. While circumstellar material, polluted white dwarfs, and the material around white dwarfs have been covered extensively in other reviews \citep[e.g.,][]{2008ARA&A..46..339W,Strom2020,2024RvMG...90..141V}, this work focuses specifically on the distinct exocometary bodies themselves rather than the source population (i.e. debris discs). It is crucial to differentiate between the cometary bodies and the circumstellar material they produce, as this distinction underlies our understanding of their activity and detectability \citep[see][for more details]{Iglesias2025}. This review does not include interstellar objects in its scope. Unlike exocomets, interstellar objects are macroscopic planetesimals originating from other stellar systems, passing through the solar system along hyperbolic paths. To date, \rev{three} such objects have been identified: largely inactive 1I/‘Oumuamua in 2017 \citep{Meech2017, Williams2017} and actively outgassing 2I/Borisov in 2019 \citep[e.g.,][]{Fitzsimmons2019,Cordiner2020}, and \rev{active 3I/ATLAS in 2025 \citep{2025ApJ...989L..36S}. These objects} hold great promise for our future understanding of extrasolar planetesimals and exocomets \citep{2023ARA&A..61..197J, 2024come.book..731F}.

\rev{Exocometary material may also be detected through observations of gas and dust released from planetesimals in collisions or other disruptive or erosional processes \citep[e.g.][]{2016MNRAS.461..845K}. Whether the parent bodies responsible for such material qualify as exocomets depends on whether their observed activity can be attributed to sublimation-driven processes, as opposed to collisional or dynamical disruption, following the definition proposed by \citet{Iglesias2025}. For instance, some studies focus on the composition of planetesimals by analysing gas and dust released in collisions or activity events \citep[e.g.][]{2017MNRAS.469S.755B}, providing indirect insights into the nature of exocometary material. Other investigations address exocomets more directly, exploring the dynamical processes that drive small bodies to short periastron distances \citep[e.g.][]{Kennedy2019} or their thermal evolution along eccentric orbits \citep[e.g.][]{Beust1990}. \re{Together, these complementary approaches contribute to a broader understanding of exocomet populations, their origins, and and how cometary material traces and participates in the evolution of planetary systems.} With these distinctions in mind, this review concentrates on the observational detection of active small bodies, exocomets, in both photometric and spectroscopic data, and on how such observations can be used to infer their occurrence, composition, and dynamical role within planetary systems.}

The detection of \rev{exocometary bodies} is largely facilitated by two main methods: photometry and spectroscopy. In photometric observations, exocomets are identified through asymmetric broadband transit light curves, which are thought to be caused by the dust coma and tail passing in front of the host star, temporarily dimming its light \citep{Lecavelier1999,Rappaport2018}. \rev{Because the absorption is caused by dust, it can in principle be wavelength dependent, producing a reddening effect where shorter wavelengths are absorbed or scattered more efficiently than longer ones. Measuring such colour dependence provides a diagnostic of dust composition and grain size.} This approach is effective in studying what is assumed to be the dusty components of the comet. Spectroscopic observations, on the other hand, focus on the gaseous atomic components of the exocomet \citep[e.g.][]{1987A&A...173..289L,1996A&A...309..474G,1997ApJ...483..449G}, through absorption lines in the stellar spectrum that change over time as the cometary material moves along its orbit. \rev{While the technique itself has been established for decades, detecting such features remains challenging because they are typically weak, transient, and can be difficult to distinguish from intrinsic stellar variability or other circumstellar gas signatures. Nevertheless, when successfully identified, these observations provide valuable information about the composition, structure, and orbital behaviour of the exocometary material.}

\rev{One important aspect of cometary activity, both in our Solar System and presumably for exocomets, is that while comet nuclei in the Solar System can sometimes be observed directly even when inactive, exocomets are currently detectable only during phases of activity.} The activity is triggered as \rev{comets} approach their host star, causing the release of gas and dust, which forms a tail and a coma. The drivers of cometary activity and the composition of the coma are heavily dependent on the distance of the comet from its host star. \rev{For Solar System comets, distant comae ($\geq$ 5 au) are typically dominated by dust and by hypervolatile species such as CO and CO$_2$ \citep[e.g.,][]{Meech2009,Jewitt2017}, as recently confirmed by detections of CO-driven activity in comet C/2014 UN271 (Bernardinelli–Bernstein) at 16.6 au \citep{2025ApJ...986L..22R} and CO$_2$-driven activity in comet C/2024 E1 (Wierzchos) at 7 au \citep{2025MNRAS.541L...8S}. Closer to the Sun, H$_2$O sublimation becomes the main driver of activity, while at very small heliocentric distances, enhanced solar insolation leads to increased atomic and ionised species \citep{Zhang2023} and volatile release through thermal degradation of grains and macromolecular material \citep{DelloRusso2016}. \re{However, while CO, CO$_2$, and H$_2$O are generally assumed to be the primary drivers of sublimation in exocomets, as for Solar System comets, this remains uncertain.} Understanding which materials are likely present in a cometary or exocometary coma at a given stellocentric distance is crucial for correctly interpreting observational signatures and inferring the composition of these bodies. During a transit, exocomets need to be close enough to the host star for signs of activity to be detectable, making their observation highly dependent on orbital geometry and proximity to the star. Because of this, only a small fraction of exocomets with favorable orbital alignments can be observed. As a result, deriving true occurrence rates or population statistics is challenging and has not yet been attempted in detail. Most studies to date therefore focus on individual detections rather than on population-level analyses.} 

Exocomets are of particular interest because they can help us understand the environments and processes that contribute to the formation and evolution of planetary systems. They likely originate from the same primordial material that formed planets and smaller bodies, and their observation can reveal the diversity of planetesimals produced at different distances from their host stars and in discs with varying compositions and structures. Studying their activity and dynamics can also provide clues about the distribution of volatiles, the presence of debris discs, and the dynamical processes that influence small-body populations, including the migration of massive planets and external perturbations such as Galactic tides or stellar fly-bys that can reshape comet reservoirs and trigger \re{enhanced comet delivery toward the inner system, sometimes described as multiple falling evaporating bodies \citep{Beust1996}}. Moreover, the study of exocomet populations around different types of stars, including A-type stars like $\beta$\,Pictoris, allows researchers to explore how stellar characteristics influence cometary activity and stability.

Despite the progress made in detecting and studying exocomets, many open questions remain. The physical properties of \rev{their} dust and gas, their frequency of occurrence in planetary systems, and \rev{their} similarities \rev{and differences with} Solar System comets are still not well understood. For example, aside from the unusual system KIC~8462852 \citep{2018ApJ...853L...8B}, there is not yet colour-dependent evidence that confirms that the photometric transits are caused by dust. Crucially, only a few tens of exocomet candidate systems have been identified, and none are on a par with $\beta$\,Pictoris in terms of the techniques that can be applied. 

\rev{In this work, we review the current state of observational exocomet research, with a particular focus on detection methods and their application to both individual systems and large-scale searches. We summarise the progress achieved through photometric and spectroscopic techniques, discuss the challenges inherent in interpreting these observations, and highlight key results from recent searches. By consolidating these observational efforts, this review aims to provide a comprehensive overview of the detection landscape and to identify promising directions for future studies that will further our understanding of active small bodies in planetary systems.}

\begin{figure} [h!]
\centering
\includegraphics[width=\textwidth]{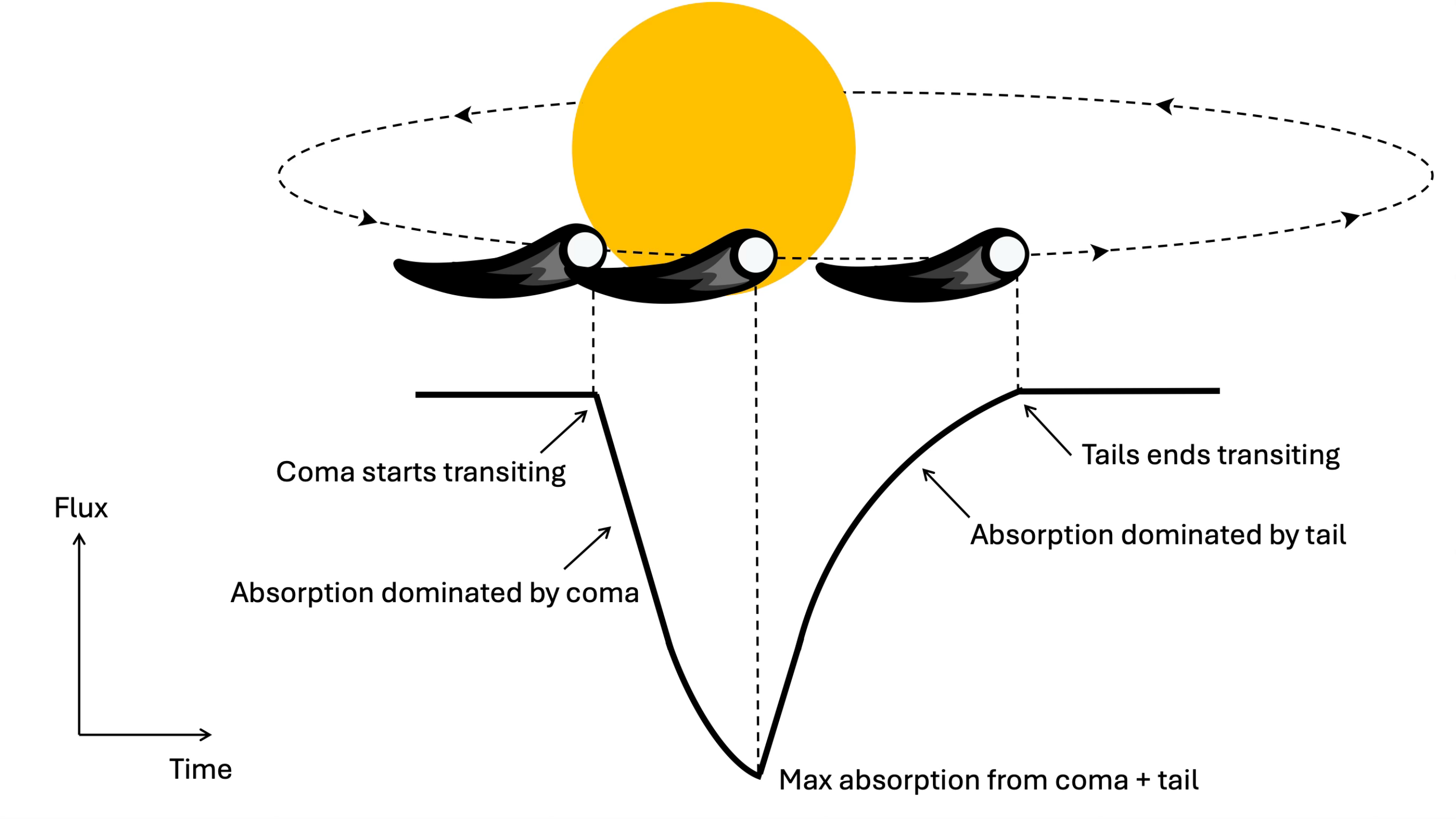}
\caption{Illustration of an exocomet transit.}
\label{fig:exocomet transit cartoon}
\end{figure}

\begin{figure}
\centering
\includegraphics[width=0.9\textwidth]{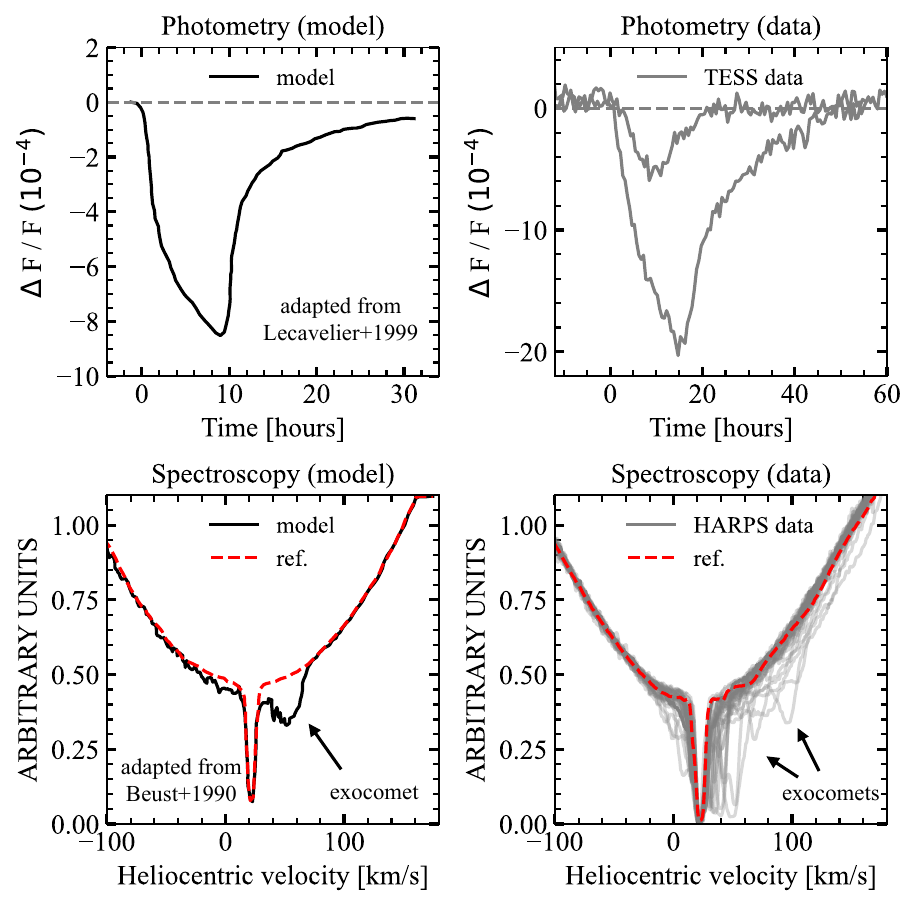}
\caption{Examples of exocomet transit detections using photometry (top) and spectroscopy (bottom). The lower plots also show deep narrow absorption from stable circumstellar gas, which indicates the systemic velocity (20\,km\,s$^{-1}$). The plots on the left are adapted from \citet{Lecavelier1999} and \citet{Beust1990}, showing the model predictions for both observation techniques. The plots on the right show exocomet detections using real data, agreeing with the model predictions. Top-right: grey lines are superimposed Transiting Exoplanet Survey Satellite \citep[\tess;][]{Ricker2015} data showing two exocomet detections in photometry with different depths from \citet{LecavalierDesEtangs2022}. Bottom-right: \rev{the plot is adapted from \citet{2014Natur.514..462K} showing superimposed High Accuracy Radial velocity Planet Searcher \citep[HARPS;][]{2003Msngr.114...20M} spectra focused on the Ca\,{\sc ii} K line in grey and a reference spectrum i.e. free of any exocomet features in red}. Any variable absorption features are attributed to exocomet transits.}
\label{fig:detection-techniques}
\end{figure}

\section{Observation Techniques}\label{sec2}

\rev{This section reviews the techniques used to detect individual exocomets transiting a parent stars.} The general picture is given by Fig.~\ref{fig:exocomet transit cartoon}; an orbiting planetesimal \rev{with an orbit inclined relative to the plane of the sky} will transit the parent star from our perspective, blocking some of the stellar flux. It is expected that the cometary nucleus itself is far too small to be detected, and it is the coma and tail that provide sufficient cross-sectional area. 

Pictured as a transit in broadband light, the \rev{event} is similar to a \rev{planetary} transit, but \rev{differs in key ways: exocomet transits are typically asymmetric due to the presence and orientation of the dust tail, which extends away from the nucleus and can be optically thin, spatially extended relative to the stellar disc, and composed of small dust grains. This geometric requirement of course implies that as for transiting planets, most systems with exocomets will not be detected because their exocomets do not transit}. 

\rev{In spectroscopy, the underlying geometry of the transit is the same as in photometry, but the signal manifests as variable absorption in specific atomic lines rather than as a broadband flux decrease. The available data are typically not of sufficient quality or duration to create a spectroscopic time-series. Asymmetric time series from the spectroscopic observations are, however, not necessarily expected} since the dynamics of atoms/ions are not necessarily the same as for dust \citep{Beust1990}, though a wavelength-based time asymmetry could be introduced by the Rossiter-McLaughlin effect \citep{1893AstAp..12..646H,1994A&A...282..804B}. 

\rev{The typical observational features of photometric and spectroscopic exocomet transit detections are illustrated in} Fig.~\ref{fig:detection-techniques}. Transits in photometry are presented as light curves (top row), while spectroscopic detections are presented by showing that one or more of multiple spectra show some variation relative to some reference where no comets are thought to be present (bottom row).

%\rev{To date, there has been no simultaneous detection of an exocomet transit in both photometry and spectroscopy. This likely reflects that the two techniques are sensitive to different components and regions of the exocometary environment: photometric transits probe the distribution of dust grains that obscure the stellar light, while spectroscopic detections trace gaseous atomic species along the line of sight. The lack of overlap may therefore be partly due to observational limitations, as simultaneous photometric and spectroscopic monitoring of the same system is rare. In addition, it may also reflect the physics of where and when these components are present, since dust can sublimate or dissipate before the gas is observed, while the gas itself may become ionised closer to the star. This suggests that photometric and spectroscopic methods intrinsically sample exocomets at different stellocentric distances.}

\subsection{Photometry}\label{photo}

Exocomets have been detected transiting stars via photometry, where an object orbiting a star causes a measurable decrease in the star's brightness as it passes between the observer and the star \citep{Rappaport2018,Kennedy2019,Zieba2019}. This absorption is broadband and expected to be caused by dust particles, so is analogous to, and thought to behave like, the dust tail of a Solar system comet. %However, only in the unusual case of KIC~8462852 has any reddening been detected, though the exocomet interpretation in that case \citep[e.g.][]{2018MNRAS.473.5286W} \rev{remains debated given the complexity of the transits and the variety of proposed periodicities ranging from $\sim$930 days to $\sim$1570 days \citep{2017A&A...608A.132K,2018ApJ...853L...8B,2018JAVSO..46...14S,2018MNRAS.473L..21B}}. 

A key characteristic used in identifying exocomets is the asymmetry of their transit light curves, caused by the cometary tail only crossing the line of sight after the coma (Fig. \ref{fig:exocomet transit cartoon}). As an exocomet transits its host star, the stellar flux decreases sharply due to the coma, the cloud of gas and dust surrounding the comet's nucleus, entering the line of sight. This is followed by an exponential increase back to full flux levels as the optically thinner dust tail continues to obscure the star \citep{Lecavelier1999}. 

% Transits with more irregular shapes have also been observed, resulting from the angle between the orbital plane and the line of sight \citep{LecavalierDesEtangs2022}. This angle critically defines the probability of observing a transit. For a transit to be detected, the exocomet's orbital plane must be sufficiently close to the observer's line of sight—nearly edge-on.

The shape of an exocomet’s transit depends on several factors, including the properties of the host star, \rev{the comet’s orbital geometry and size, and the effects of radiation pressure, which influence the morphology and evolution of the dust tail and coma \citep{2024arXiv241006248S}}. Stellar winds may also influence the evolution of cometary tails, but at the time of writing there is no observational evidence of its effects on exocomets (see \cite{Vrignaud2025}). Importantly, symmetric exocomet transits are also possible when the comet dust tail is aligned along the line of sight because the cometary motion during transit has a significant radial component. \rev{This typically occurs for eccentric orbits whose periastron lies roughly perpendicular to the observer’s line of sight, as illustrated in the symmetric case of \citet[their Fig. 2]{Lecavelier1999}.} These symmetric transits are nearly indistinguishable from symmetric transits caused by binary stars or exoplanets. Therefore, the asymmetry in the transit profile is a crucial diagnostic feature for exocomet detection, but will not necessarily \rev{reveal} all transiting exocomets.

Estimating the physical parameters of exocomet bodies from transit light curves is challenging, as these events are rare and have been based only on a few observations. However, various modelling approaches have been developed to characterise some physical parameters \rev{of transiting exocomets}, both numerically and analytically \rev{\citep[see][for an overview of the methods]{2024arXiv241006248S}}. \rev{Early numerical models by \citet{Lecavelier1999} simulated photometric variations caused by dust distributions around exocomets, while more recent studies have expanded this to multi-wavelength analyses \citep{Kalman2024} and Monte Carlo techniques \citep{Lukyanyk2024}. \re{Such models generally require detailed knowledge of the host system, including stellar parameters and viewing geometry.} They typically rely on assumptions about the size and composition of dust grains, the dust production rate, the optical depth of the tail, and the orbital configuration of the comet, all of which remain uncertain and can strongly influence the modelled transit shape and depth.}

A simpler empirical alternative focuses on the key characteristic of known exocomet transits: their asymmetric profile. Empirical models have been developed to fit transit light curves, which is beneficial for initial detections before attempting detailed physical parameter estimations. For example, \citet{Zieba2019} developed a simple exponential tail model that fits well to the first $\beta$\,Pictoris detection. \citet{LecavalierDesEtangs2022} followed this up with a two-component exponential model to measure the transit depths and estimate the exocomet size distribution for the system. \citet{Kennedy2019} introduced a half-Gaussian, half-exponential model representing the comet tail, suitable for large-scale searches due to its simplicity. \citet{2025MNRAS.542.1486N} used a skewed Gaussian model, providing continuous fits to the data. These \rev{empirical} models do not yield many physical parameters, but with certain assumptions, they can yield a lower limit on the transverse orbital velocity \citep{2017RSOS....460652K,Zieba2019}, which is valuable for constraining the orbit.  %While empirical fits do not provide detailed physical information about the cometary bodies, they are valuable for identifying new candidates, which can then be characterized in detail with numerical models.

Somewhat similar asymmetric transits have been detected around white dwarfs and are interpreted as minor bodies disintegrating around these stars \citep[e.g.,][]{Vanderburg2015, Vanderbosch2020}. These detections challenge the strict definition of an exocomet, as the observed signals arise \re{primarily from dust and gas produced by tidal disruption of the bodies themselves, rather than from sublimation-driven activity of intact cometary nuclei directly observed in transit. This contrasts with other exocomet systems around main-sequence stars, where the transit signatures are interpreted as dust tails generated by enhanced sublimation from surviving remnants of disrupted bodies.} While periodicity has been established for some systems, many do not exhibit any \citep{2025PASP..137g4202B}.

%\begin{figure}
%    \centering
%    \includegraphics[width=0.99\linewidth]{LC_Fit_VI_7_2016.pdf}
%    \caption{Phase-folded light curves of WD 1145+017 showing an asymmetric transit in the UV, optical, Ks band, and 4.5 $\mu$m, respectively, observed on 2016 March 28–29. The black line represents the best-fitting
%models \citep{Xu2018a,Xu2019a}.}
%    \label{fig:wds}
%\end{figure}

\subsubsection{Searching for Exocomets in Photometry} 

\rev{Searches for exocomets in photometry have been conducted with two main goals: long-term monitoring of specific systems known from spectroscopy (most notably  $\beta$\,Pictoris), and wide-field transit surveys aimed to build their demographics and statistics. Both approaches face the challenge of identifying the very shallow, asymmetric signals characteristic of exocomet transits.}

\rev{The first dedicated photometric searches were conducted for $\beta$\,Pictoris between 1975 and 1992 \citep{1995A&A...299..557L,1997A&A...328..311L,1997A&A...328..321L}, which included a dimming event in 1981 later interpreted as evidence of exocometary activity. These efforts represented the first targeted attempt to detect photometric variability consistent with comet-like behaviour around another star. However, typical ground-based precision at the time ($\sim$0.3-1\%) was insufficient to detect the shallow $\sim$0.1\% asymmetric transits predicted by models \citet{Lecavelier1999}.}

\rev{Decades later, the advent of high-precision space missions dedicated to exoplanet transit surveys such as the COnvection, ROtation and planetary Transits \citep[\Corot;][]{2009A&A...506..411A}, \kepler \citep{2010Sci...327..977B}, \ktwo \citep{2014PASP..126..398H}, and Transiting Exoplanet Survey Satellite \citep[\tess;][]{Ricker2015} revolutionised this field and motivated large-scale searches for exocomets. The first detections of likely exocomet transits around KIC~3542116 and KIC~11048727 were discovered in the \kepler data by manual visual inspection of light curves \citep{Rappaport2018}. Prior studies of KIC~8462852 (Tabby’s Star) had also proposed that families or strings of comets could explain its irregular dimming behaviour \citep{Boyajian2016,2016ApJ...819L..34B,2017A&A...608A.132K}. These works suggested that the breakup of larger bodies could reproduce the observed asymmetric and multi-structured transit patterns, providing some of the first tentative photometric evidence for exocomets beyond $\beta$\,Pictoris.}

\rev{Building on these initial findings, automated approaches were then developed to process larger datasets efficiently, reduce human error, and increase efficiency in processing the large data volumes \citep{Kennedy2019}. Automated approaches also allow for the exploration of different hypotheses that would be difficult to implement manually, such as testing different detrending methods, empirical comet models, and applying different parameter thresholds to narrow down candidates. These automated searches, adapted from exoplanet transit detection, focus on single, non-periodic events. As exocomets are unlikely to transit repeatedly within a mission’s lifetime, the detection criterion is based mainly on transit shape.  The asymmetry of an exocomet transit is typically quantified by fitting simple empirical models such as skewed or half-Gaussian profiles \citep{Kennedy2019,2025MNRAS.542.1486N}. These models are computationally efficient compared to numerical models while preserving the essential transit morphology.}

\rev{More recent studies have explored machine-learning (ML) methods, such as Random Forest classifiers \citep{Dobrycheva2023} and Neural Networks \citep{Dumond2025}, trained on synthetic exocomet light curves to distinguish comet-like events from false positives due to the rarity of real photometric detections. As datasets continue to grow with new \tess sectors and upcoming missions like the PLAnetary Transits and Oscillations of stars \citep[\plato;][]{2025ExA....59...26R} mission and the Nancy Grace Roman Space Telescope \citep[Roman;][]{Spergel2015,Akeson2019}, ML approaches will be increasingly useful for identifying diverse and rare transit morphologies. Nevertheless, all automated searches produce candidate lists that require careful manual vetting.}

\rev{A key step in any photometric search is removing intrinsic stellar variability that can obscure shallow cometary transits. Many stars display periodic or quasi-periodic variations, such as $\delta$ Scuti pulsations, that must be modelled and subtracted. For instance, $\beta$\,Pictoris, exhibits $\delta$ Scuti-like oscillations that were removed to clean the light curves, revealing the exocomet transits seen in \citet{Zieba2019, Pavlenko2022, LecavalierDesEtangs2022}. Similar detrending has been applied to CHaracterising ExOPlanet Satellite \citep[\cheops;][]{2021ExA....51..109B} observations of HD~172555 \citep{Kiefer2023}. For large-scale searches, individual modelling is impractical, and more generalised detrending is favoured using frequency-based filtering such as Lomb–Scargle periodograms \citep{Kennedy2019} or median-based time-domain methods \citep{2025MNRAS.542.1486N}.These techniques efficiently remove stellar signals while preserving the asymmetric transit shapes characteristic of exocomets.}

\rev{Large photometric datasets inevitably contain instrumental and astrophysical sources of false positives. Step-like instrumental systematics in \kepler and \tess light curves can mimic asymmetric transits after detrending, and have to be accounted for in these search pipelines. Astrophysical false positives include rotational modulation, young stars with circumstellar discs showing irregular ``dipper’’ behaviour \citep{Ansdell2016,Ansdell2019,Hedges2018}, disintegrating planets \citep{Brogi2012,Rappaport2012,Hon2025}, and heartbeat binaries showing periastron-related variability \citep{2012ApJ...753...86T}. As a result, distinguishing true exocomets from these false positives requires careful inspection of the candidates and their host stars.}

\rev{White dwarfs offer particularly stable photometric conditions for detecting disintegrating material, as most non-pulsating single white dwarfs vary by less than 1\% \citep{Hermes2017}. Despite their faintness and short transit durations, more than a dozen systems now exhibit transiting or disintegrating debris, identifiable through characteristic von Neumann statistics and Pearson skewness signatures \citep{Vanderburg2015,Vanderbosch2020,Vanderbosch2021,Farihi2022, 2025PASP..137g4202B,2025ApJ...980...56H}.}

\subsection{Spectroscopy}\label{spec}

\rev{As noted earlier, spectroscopic transits share the same geometric principle as photometric ones, with material passing in front of the stellar disc. However, rather than broad-band dimming, the signal manifests as transient, typically Doppler-shifted absorption lines produced by gas in the exocomet’s coma \citep{Ferlet1987, 1987A&A...173..289L}. By analogy with Solar System comets, the observed species trace the coma and ion tail of the exocomet. The ion-tail analogy is particularly relevant, as the ion tail of a comet points away from the Sun, and it is likely that the same geometry applies to exocomets \citep{Beust_Tagger_1993}. In the case of $\beta$\,Pictoris, radiation pressure rather than a stellar wind is thought to be the dominant force driving the ions away from the star. Because spectroscopic detections trace gas absorption rather than dust extinction, they do not necessarily produce asymmetric transit profiles, and thus transit asymmetry is not a defining criterion for identifying spectroscopic exocomet events. In any case, for $\beta$\,Pictoris it is thought that the absorption is primarily from material in the coma of the comet, as the ions in the tail are quickly driven to high velocity, and consequently, the tail is much less dense than the coma and harder to detect \citep{Beust1989, Beust1990, Beust1991_caii_modelling}}.

\rev{When the cometary coma transits the star, additional absorption features appear superimposed on the stellar spectrum. As the comet moves radially relative to the star, these lines are Doppler-shifted according to the velocity difference between the gas and the stellar rest frame \citep{Beust1990, Beust_Tagger_1993}. A cometary body moving toward the star absorbs stellar photons that are slightly blueshifted compared to their absorption wavelength in the stellar frame; therefore, the resulting feature appears redshifted in the stellar spectrum. Consequently, a redshifted absorption feature indicates material moving toward the star and away from the observer, while a blueshifted feature corresponds to material moving toward the observer and away from the star, consistent with the conventional Doppler interpretation of red and blue shifts. These shifts provide direct information about the kinematics of the transiting gas and the mechanisms governing its motion \citep{Beust1991_planet,Beust_Tagger_1993, Beust_Morbidelli_1996, Beust2024}.}

\rev{Sporadic, short-lived absorption features are the hallmark of exocomet detections. Because the gas is only observed while transiting in front of the star, features vary in both depth and velocity over timescales of minutes to hours, sometimes appearing to accelerate during transit \citep{Kennedy2018}. The duration of an absorption feature depends on the gas velocity: faster-moving material produces briefer events, while slower gas yields longer-lived signatures \citep{Lagrange1996, Beust1998}. For $\beta$\,Pictoris, for instance, absorptions corresponding to velocities of 30~kms$^{-1}$ or 300~kms$^{-1}$ are expected to last roughly 6 hours or 30 min, respectively \citep{Beust1996}. In practice, complete transits are rarely observed, except in intensive monitoring campaigns such as those of  $\beta$\,Pictoris, in the 1990s \citep[e.g.,][]{Lagrange1988, Lagrange1992, deleuil1993, vidal-madjar1994} and radial velocity planet search campaign \citep{2014Natur.514..462K,2019NatAs...3.1135L}, where hundreds of spectra per night were taken.}

\rev{A variety of ions have been detected in spectroscopic exocomet transits, with the main events typically identified in the Ca\,{\sc ii} H \& K lines. Ultraviolet (UV) spectroscopy has also revealed additional variability around species with higher ionisation potentials such as Fe\,{\sc ii}, Mg\,{\sc ii}, and C\,{\sc ii}, offering a complementary probe of the physical and chemical conditions in exocometary gas.}

\rev{\subsubsection{Searching for Exocomets in Spectroscopy}\label{specsearch}}

\rev{Spectroscopic searches identify exocomets by detecting transient absorption features that vary between observations. Compared with photometry, the spectral dimension is rich, while the time coverage is sparse. Thus, searches are performed by comparing spectra as illustrated in the bottom row of Fig.~\ref{fig:detection-techniques}. Common to all spectroscopic searches, a reference spectrum, free of any exocometary features, is needed to help identify transient absorption features that could be potential exocomet detections. This reference spectrum consists of the stellar photosphere combined with stable spectroscopic features \citep[e.g., interstellar gas or circumstellar gas absorption,][]{Hobbs1985, Iglesias2018, Rebollido2020}. In practice, multiple spectra are essential, since an accurate reference spectrum should not capture any variable features. For most stars, this reference is obtained as the median of all spectra, though for $\beta$\,Pictoris, a tailored approach is necessary because no spectrum is entirely free of exocomet absorption \citep{2014Natur.514..462K}.}

\rev{Two broad search strategies exist. The first involves detailed inspection of individual targets such as for $\beta$\,Pictoris,  HD\,172555 and 49\,Ceti \citep{2014A&A...561L..10K,2016ApJ...824..126M} or small samples of stars \citep[e.g.,][]{Lecavelier1997d, 1997A&A...325..228L, Montgomery2012, Iglesias2018,Rebollido2018}, while the second uses semi-automated methods applied to many targets \citep{2025MNRAS.537..229B}. In all cases, genuine exocometary events are expected to manifest as transient, often weak absorption features, either red- or blueshifted, with variable widths and depths. Ideally, such features appear coherently in multiple lines (e.g. \re{Ca\,{\sc ii}, Na\,{\sc i}, }Fe\,{\sc ii}, Mg\,{\sc ii}, and C\,{\sc ii}) with variable depths depending on differences in oscillator strengths and optical thickness of the gas \citep{Beust1990, Lagrange1992, 1997A&A...325..228L, 2014Natur.514..462K}. The signal-to-noise ratio in spectra can sometimes be poor, which is why quantifying detection significance against the background noise is essential. \citet{2025MNRAS.537..229B} proposed an attempt at a formalised approach that estimates spectral noise and the uncertainty on the reference spectrum, providing the first step toward standardised detection metrics. Although less sensitive to weaker variations that are easier for experts to identify by-eye analysis, such quantification enables reproducible, large-scale searches. Future improvements will likely involve more sophisticated modelling of exocomet-free reference spectra, better characterisation of stellar noise, and the application of machine-learning algorithms to enhance sensitivity and reduce false positives.}

\rev{Spectroscopic detection is complicated by telluric absorption from Earth's atmosphere, which can mimic or obscure genuine exocometary features. Observations, therefore, focus on spectral regions less affected by tellurics. Ultraviolet observations from space-based instruments such as the \textit{Hubble Space Telescope} have revealed strong absorption from metal ions (e.g. Fe\,{\sc ii}, Mg\,{\sc ii}, and C\,{\sc ii}), where many of the first FEB-related absorptions were detected \citep{Lagrange1986, 1987A&A...173..289L}. From the ground, most searches target the Ca\,{\sc ii} H\&K lines, which are less impacted by telluric absorption and historically the main diagnostic of Falling Evaporating Bodies \citep{Ferlet1987, Lagrange1992, 2014A&A...561L..10K}. Other lines such as Na\,{\sc i} have been used successfully \citep[e.g.,][]{2014Natur.514..462K,Iglesias2018,Rebollido2020,2025A&A...700A.239H}, though careful modelling of telluric contamination is essential.}

\begin{figure}
\centering
\includegraphics[page=2, width=0.32\textwidth]{HIP61782_CaK_Res.pdf}
\includegraphics[page=1, width=0.32\textwidth]{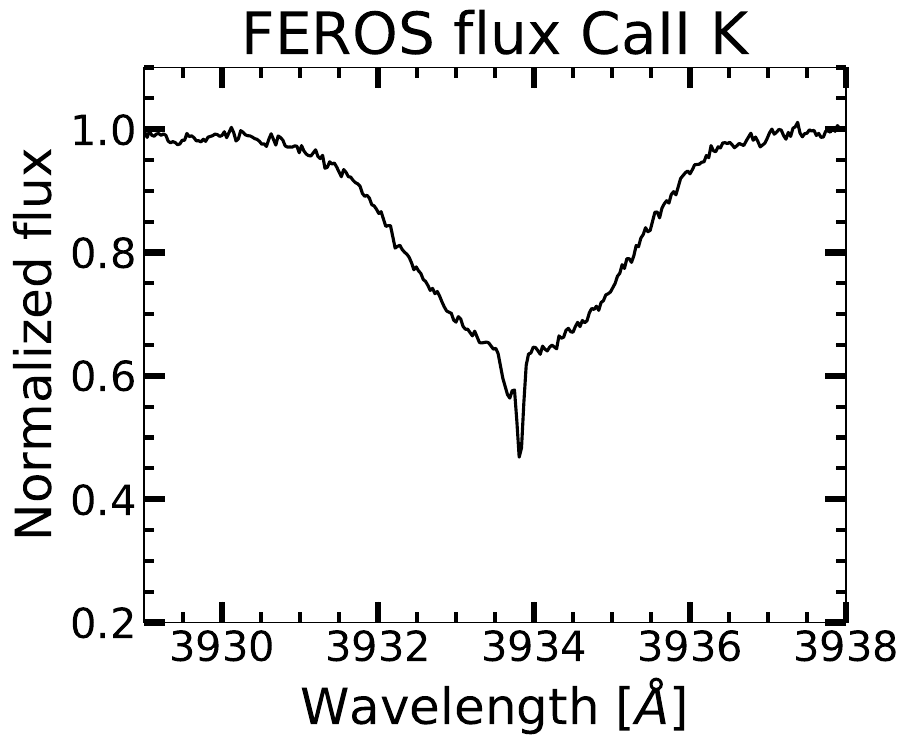}
\includegraphics[page=3, width=0.32\textwidth]{HIP61782_CaK_Res.pdf}
\caption{Example of gas absorptions in HD\,110058 observed under different resolutions. The spectrum shows the wide \rev{rotationally broadened stellar absorption} signature, a narrow and deep circumstellar disc absorption, and a slightly wider \rev{interstellar} absorption just blueward of the circumstellar disc line.  Left: HARPS observation, $R\approx115000$, absorptions are resolved. Middle: FEROS observation, $R\approx48000$, absorptions are fairly resolved and detected at lower intensity. Right: X-Shooter observation, $R\approx5500$, absorptions are no longer resolved and detection is difficult. }
\label{fig:resolution}
\end{figure}

\rev{High spectral resolution is crucial to isolate exocometary absorption from circumstellar and interstellar features.} Instruments such as the High Accuracy Radial velocity Planet Searcher \citep[HARPS; $\lambda$\,$\in$\,(380--690)\,nm, $R\approx110\,000$, $\approx3$\,\kms;][]{2003Msngr.114...20M} can achieve signal-to-noise ratio (SNR) $>$250:1, allowing subtle variations to be detected  (see left panel in Fig.~\ref{fig:resolution}). Medium-resolution instruments like EsPaDOnS \citep[$\lambda$\,$\in$\,(370--1050)\,nm, $R\approx65,000$, $\approx4.6$\,\kms;][]{2006ASPC..358..362D} can achieve typical SNR of around 200:1 \citep{Welsh2015}. The Fiber-fed Extended Range Optical Spectrograph \citep[FEROS; $\lambda$\,$\in$\,(360--920)\,nm, $R\approx48,000$, $\approx6.2$\,\kms;][]{1999Msngr..95....8K}, with a lower resolution, is still sufficient to detect exocomets, though \rev{absorptions are now blended} (see the middle panel in Fig.~\ref{fig:resolution}). In contrast, low/medium-resolution measurements such as X-shooter \citep[$R\approx5500$, $\sim$50\,km\,s$^{-1}$,][see the right panel in Fig.~\ref{fig:resolution}]{2011A&A...536A.105V} lack the sensitivity to detect absorption features additional to the main stellar spectrum, making it challenging to identify deep circumstellar disc absorption or any more subtle variations produced by exocomets.

\rev{As with photometric searches, candidate detections require careful vetting for false positives. Instrumental artefacts, including detector response variations, wavelength-calibration drifts, and imperfect continuum normalisation, can produce or hide spurious signals. Astrophysical sources of confusion include variable stellar activity, pulsations, binarity, and quasi-stable circumstellar or interstellar absorption that may appear transient due to noise or sparse sampling. Notable false positives include HR~10, originally interpreted as hosting exocomets \citep{1990A&A...227L..13L,1997A&A...325..228L} but later shown to be a binary system with circumstellar envelopes \citep{2019A&A...629A..19M}, and $\phi$\,Leo, where the variability was traced to $\delta$\,Scuti pulsations \citep{Eiroa2021}.  
Most ground-based searches have so far focused on the Ca\,{\sc ii} doublet, although for late-type stars this region can vary strongly due to stellar magnetic activity. Such perturbations are easily identified as emission features but difficult to remove, reducing sensitivity to absorption features and limiting confident exocomet detection around active stars.}

\section{Observational Results}\label{occurence}

\rev{This section summarises the observational outcomes of exocomet searches using the techniques described in Section~\ref{sec2}. Although the detection principles are fundamentally similar, identifying transient absorption in spectroscopy (Section~\ref{spec}) or asymmetric broadband dimming in photometry (Section~\ref{photo}), the practical results obtained so far differ substantially between the two methods.} 

\rev{To date, there has been no exocomet transit that has been detected in both spectroscopy and photometry simultaneously. This absence of joint detections likely reflects both observational and physical factors: spectroscopic and photometric monitoring campaigns are rarely coordinated to observe the same target simultaneously, and as discussed in Section 2, the two techniques may probe different radial distances that favour either gas ionisation (for spectroscopy) or dust production (for photometry). Identifying and characterising exocomets with multi-method techniques is therefore a key future goal, and we return to this topic in Section 4.}
%This section discusses the results of dedicated exocomet searches in both spectroscopy and photometry. We discuss the outcomes of those studies and the insights from their work. %\rev{In this context, surveys refer to large spectroscopic or photometric datasets (e.g., HARPS, \textit{Kepler}, TESS) that have been analysed for exocomets, rather than to dedicated observing programs.}

\subsection{Spectroscopy}

Detections in spectroscopy have a long history, preceding the detection of exoplanets \citep{Ferlet1987}. These have generally followed the methods outlined \rev{in Section \ref{spec}}, and the two main wavelength regimes \rev{used} are \rev{the} optical, primarily the Ca\,{\sc ii} K line at 3933.7 \AA, and {the} UV, which probes heavier atoms such as Fe, Al, Mg. The \rev{first detections} prompted theoretical efforts to understand the physics of the coma and tail \citep[e.g][]{Beust1989, Beust1990, Beust1991_caii_modelling, Beust_Tagger_1993}, which showed that a ``falling evaporating body'' (i.e. an exocomet) was consistent with the observed signatures. 

\rev{Since the first discovery of exocomets in $\beta$\,Pictoris, many spectroscopic searches have followed, looking for analogues around other stars \citep[e.g.][]{Lagrange-Henri1990b,Lecavelier1997d,Welsh1998}. Initial efforts concentrated on a few ``shell stars'', objects suspected to host circumstellar gaseous envelopes.} Later, searches expanded to stars with debris discs, $\lambda$ Bootis stars, and stars with previous detections of circumstellar gas at mm wavelengths \citep[e.g.][]{Montgomery2012,Iglesias2018,Rebollido2018,Rebollido2020}. Most of these searches focused on A-type stars, \rev{a bias that likely arose from the early success with $\beta$\,Pictoris.} \rev{A-type stars are not only younger and more likely to host detectable debris discs than later-type stars, but their rapid rotation also broadens stellar absorption lines, producing a smoother continuum that facilitates the identification of narrow exocometary absorption features.} 

\rev{More recently, the spectral-type coverage has been expanded using the HARPS archive \citep{2025MNRAS.537..229B}, which is dominated by FGKM stars as they are relatively easier targets for detecting planets through radial velocity methods. Although this broadened the target sample, no spectroscopic exocomet detections have yet been identified among these later-type stars. Since there is clear evidence for photometric exocomet detections in F-type stars \citep{Rappaport2018,Kennedy2019}, and given that our own G-type Sun hosts comets, \rev{the lack of spectroscopic detections among FGK stars is very likely the result of observational bias rather than a true absence of exocomets}. As described in Section~\ref{spec}, detecting exocomet absorption in the Ca\,{\sc ii} doublet becomes increasingly difficult for later-type stars because chromospheric activity introduces variable emission or absorption in the same lines, strongly masking or mimicking the narrow, transient features expected from exocomets. In addition, the detection of exocometary gas absorption requires very high SNR ($\gtrsim 100$), which is more difficult to obtain for intrinsically fainter FGKM stars within reasonable exposure times. Ultraviolet spectroscopy, which is less affected by these issues and sensitive to multiple atomic species, remains sparse due to cost and limited availability, so only a small number of stars have been observed in the UV.} %\rev{because data in the UV} are more scarce and expensive, but with the advantage of sensitivity to multiple atomic species. A relatively small number of stars have consequently been observed in the UV.

\rev{Nevertheless, individual detections outside $\beta$\,Pictoris have been reported, including HD~172555 \citep{2014A&A...561L..10K,Kiefer2023}, and 49~Ceti \citep{2016ApJ...824..126M}. Because exocomet transits are intrinsically rare, most published spectroscopic detections originate from searches intentionally targeting stars already known to host circumstellar gas such as shell stars, debris-disc systems, and stars with millimetre detections of gas \citep[e.g.,][]{1990A&A...227L..13L,1997A&A...325..228L, Montgomery2012, Welsh2015, Eiroa2016, Welsh2018, Rebollido2020}.} For a complete list of detections in spectroscopy, see Table~1 in \citet{Iglesias2025}.

\rev{Although $\beta$\,Pictoris provides the only system where detailed compositional information has been obtained for multiple exocomets, virtually nothing is known about the chemical composition of exocomets beyond this archetypal case. While multiple species have been detected for $\beta$\,Pictoris, and a few for 49~Ceti and 51~Oph \citep[][respectively]{2016ApJ...824..126M,2002ApJ...568..343R}, most systems are either detected only in Ca\,{\sc ii} or broadband photometry, providing little compositional information. The planetesimals that probably represent the source regions of exocomets, observed as debris discs, are thought to outgas CO \citep[e.g.][]{1995Natur.373..494Z,2011ApJ...740L...7M,2014Sci...343.1490D}, but no other cometary species have yet been detected (aside from the atomic products C and O). This stands in stark contrast to metal-polluted white dwarfs, where in many cases only composition is known, yet the dynamical context is largely inaccessible.}

\rev{Recent efforts to analyse archival UV spectra of $\beta$\,Pictoris’ exocomets have nevertheless begun to open a path forward. Studies by \citet{2024A&A...684A.210V} and \citet{2024A&A...691A...2V} have revealed Solar-like abundances of Ni+/Fe+ and Cr+/Fe+ in these exocomets. Extending similar analysis to other systems, like 49~Ceti or new candidate systems, could broaden our understanding of exocomet compositions beyond $\beta$\,Pictoris and placing exocomet compositions into a broader comparative context.}

\rev{We refer the reader to \cite{Vrignaud2025} for a detailed review of the processes governing ion populations in exocometary comae. A key challenge for future simulation efforts is determining the density and temperature profiles expected for exocomets. Promising progress has been made by \cite{2024A&A...684A.210V}, who demonstrated how Fe\,{\sc ii} column densities in exocomets around $\beta$\,Pictoris can be used to constrain these conditions.}

\subsection{Photometry}

Thanks to the rapid advancement in exoplanet science over the past few decades, a sizeable volume of high-precision photometric data has been searched for exocomets. \rev{As described in Section~\ref{photo}, photometric detections rely on identifying shallow, asymmetric transit signatures produced by dust in the coma and tail. Ground-based precision is so far still insufficient to detect the predicted $\sim$0.1\% events \citep{Lecavelier1999}, and the first confirmed photometric detections appeared with the advent of the space-based \kepler mission.} \citet{Rappaport2018} performed a visual search of \kepler light curves and identified two candidates, KIC~3542116 and KIC~11084727, likely hosting transiting exocomets. \citet{Kennedy2019} subsequently carried out an automated search and confirmed these events, adding a third candidate (KIC~8027456). All occurred around A/F-type stars, consistent with the spectroscopic detection bias described in Section~\ref{spec}. More recently, \rev{a machine learning approach across the full \kepler dataset has also uncovered} an additional 10 new candidate events \citep{Dumond2025}. The spectral-types for the candidates in this work are more varied than \cite{Rappaport2018} and \cite{Kennedy2019}, with detections around later-types, which may suggest photometric detections can probe detections around later-types easier than spectroscopic methods. However, as with all exocomet candidates, characterising these detections in detail along with multi-method observations will be necessary.

%\kepler was the first mission that provided sufficient photometric precision to enable the detection of exocomet candidates. \citealt{Rappaport2018} \rev{conducted a visual} search in the \kepler data to search for unusual transiting events, where they identified two systems, KIC~3542116 and KIC~11084727, concluded to be \rev{likely transiting exocomets}. \citealt{Kennedy2019} followed this work up with an automated search in the same dataset, where they recovered the two candidates and identified a new candidate: KIC~8027456. All of these candidates were detected around A/F-type stars, matching the spectral types for exocomet-host stars identified by spectroscopy. \rev{\citet{Kennedy2019} therefore concluded that because the \kepler exocomet detections were around A/F types that i) the detections were probably real, and ii) that A/F types appear more likely to have detectable exocomets than later types. More recently, \cite{Dumond2025} performed a search across the \kepler dataset using machine learning, identifying 10 new events.} 

\rev{The successor mission of \kepler, \tess, provided the first photometric detections of exocomets around $\beta$\,Pictoris itself \citep{Zieba2019,Pavlenko2022,LecavalierDesEtangs2022}, making it the first system to have detections in both photometry and spectroscopy (though not simultaneously). Multiple transits in its 8 \tess sectors of observation enabled the first empirical estimate of the exocomet size distribution in this system \citep{LecavalierDesEtangs2022}. \rev{The exocomet size distribution in} $\beta$\,Pictoris \rev{is consistent with that observed for comets in the Solar System, supporting physical models in which collisional fragmentation can shape the size distributions of exocomets in young planetary systems \citep{Dohnanyi1969, Obrien2005}.} A similar size distribution was inferred for exocomets transiting the extreme debris disc system RZ~Psc \citep{Gibson2025}, supporting this interpretation.}

%The successor to the \kepler mission was \tess, \rev{which also enabled the first photometric detections of exocomets around $\beta$\,Pictoris \citep{Zieba2019, Pavlenko2022,LecavalierDesEtangs2022}. These detections confirmed $\beta$\,Pictoris to be the first system with observations of exocomets in photometry and spectroscopy. The multiple photometric transit events detected around $\beta$\,Pictoris with \tess have also allowed the first estimates of the exocomet size distribution in this archetypal system to be calculated \citep{LecavalierDesEtangs2022}.} \rev{The exocomet size distribution in} $\beta$\,Pictoris \rev{is consistent with that observed for comets in the Solar System, supporting physical models in which collisional fragmentation shapes the size distributions of exocomets in young planetary systems \citep{Dohnanyi1969, Obrien2005}.} \rev{\tess has also identified likely exocomets around the extreme debris disc system RZ PSc \citep{Gibson2025}; 24 photometric detections across three sectors constrained an exocomet size distribution consistent with both Solar System comet populations and $\beta$\,Pictoris, further supporting collisional fragmentation.}

Large-scale automated photometric searches have also been attempted using \tess. \citet{2025MNRAS.542.1486N} searched the data from \tess primary’s mission and reported five new candidates: three around main-sequence stars (two A/F-type, one likely G-type) and two around giants. These results indicate that photometric exocomet detections remain rare but are not restricted solely to early-type hosts. The results for the automated searches in \kepler and \tess effectively conclude that photometric observations of exocomets are rare, and that more detections are needed to build a significant sample for exocomet demographics. \re{Other current photometric surveys include \cheops, where transits around HD 172555 provides compelling evidence as the second with detections in both spectroscopy and photometry (\citealt{Kiefer2023}).} \tess is continuing to observe the night sky, and with \plato nearing launch, future searches for exocomets with upcoming datasets are well-motivated. \rev{A  list of published photometric detections is provided in Table~2 of \citet{Iglesias2025}.}

%\rev{An automated search for exocomets in the \tess primary mission has also been conducted in \citep{2025MNRAS.542.1486N} to estimate the frequency of exocomet detection across the host star's spectral types, with the promise that an all-sky survey may yield increased detections around young stars as seen in spectroscopy. The search yielded five new candidates, along with the recovery of the exocomet detection around $\beta$\,Pictoris. Three of the candidates are main-sequence: two in the A-F spectral type range, the other more likely a G-type. The other two candidates are around giant stars, so their origin is less certain. Both large-scale surveys conclude that photometric observations of exocomets are rare, and that more detections are needed to build a significant sample for exocomet demographics. TESS is continuing to observe the night sky, and with PLATO nearing launch, future searches for exocomets with upcoming datasets are well-motivated.}

% one of the candidates may be an exocomet transit, but the other candidate host is probably a G-type supergiant, suggesting intrinsic variability as a simpler explanation.

% The search yielded an occurrence rate similar to the search with \textit{Kepler}. 

% primary mission resulted in 5 new exocomet candidates in their work, and concluded discussed the occurrence rates of exocomets in TESS. Other photometric surveys included CHEOPS, where the detection of an exocomet candidate in HD 172555 has also cemented the star as an exocomet host in both spectroscopy and photometry (\citealt{Kiefer2023}). 

It is \rev{also} important to emphasise exocomet searches that have resulted in null detections. \rev{For instance,} \citet{Rebollido2023} observed the star 5~Vul \rev{with \cheops, despite its known spectroscopic exocomet activity \citep{Montgomery2012, Rebollido2020}}, but identified no photometric transits. \rev{The detection frequency of exocomet features can therefore vary widely between spectroscopic and photometric observations. For example, $\beta$\,Pictoris shows spectroscopic transits on a daily basis, whereas photometric transits are observed far less frequently.} %One possible explanation for 5~Vul is simply that the transits are not frequent enough that there was a photometric transit during the \cheops observations. For $\beta$\,Pictoris, however, spectroscopic transits are nearly always seen. 
\rev{One explanation is that the two techniques are likely sensitive to different regions of the system. The photometric exocomet detected by \citet{Zieba2019} was inferred to orbit at roughly 3~au from the star, while spectroscopic detections typically trace gas located much closer in, at distances ranging from a few to a few tens of stellar radii, depending on the atomic species observed \citep{Beust1990, Beust1998}. These radial scales correspond to the different physical environments: at larger distances, dust can survive and produce observable detections in photometry, whereas closer to the star, high temperatures and strong radiation fields cause dust to completely sublimate and atoms to become ionised, producing variable absorption lines in spectroscopy \citep{Rebollido2023,Heller2024,Vrignaud2025}.}

% It may imply that the dust tails seen in photometry are completely sublimated by the time at which they reach the distance of spectroscopic observations, or the opposite, where the gas tails are not ionised enough at photometric distances.

\rev{Photometric surveys have their own limitations, which are mostly driven by the details of the mission in question. \kepler provided excellent precision but observed a limited stellar population dominated by later-type stars, whereas most previous spectroscopic exocomet detections were around early-type stars. \tess offers near all-sky coverage, reducing host-star bias, but its 27-day observing windows and lower per-point precision limit sensitivity to rare and shallow events, especially around faint stars. A combination of future missions with \kepler-like precision and wide-field coverage, such as \plato, Earth 2.0, and Roman, will likely be essential to build a statistically meaningful sample of photometric exocomet detections.}

\subsection{Occurrence rates}

Most \rev{exocomet searches} in the literature are highly biased \rev{to increase detection chances}, making it challenging to accurately assess exocomet occurrence rates. \rev{Spectroscopic} detections are predominantly observed around A-type stars, with a few detected around F and B-type stars.
% This being said, this trend seems to differ from exocomet detections in photometry, which could hint at observational biases in addition to a biased sample of targets.
\rev{The main source of bias is target pre-selection: previous searches focused on stars already known to exhibit spectral variability or circumstellar gas \citep[e.g.,][]{1997A&A...325..228L,Eiroa2016,Rebollido2020}, inflating detection rates to $10-20$\% \citep{Iglesias2018, Welsh2018, Rebollido2020}. However, spectroscopic observations are typically limited to the brightest targets, restricting sample sizes and skewing detection statistics.}
% Additionally, spectroscopic observations are typically limited to the brightest targets, limiting the sample to small numbers of stars.}
% Removing the biases in previous surveys to do statistics is not always trivial. We can use \citet{Iglesias2018} and 
% \citet{2025MNRAS.537..229B} to estimate the occurrence rate of exocomet transits in spectroscopy using the Ca\,{\sc ii} lines \rev{by accounting for survey completeness and detection biases in their statistical frameworks.}

\rev{Correcting for these pre-selection biases is not trivial. The ${\approx}10$\% detection rate of \citet{Iglesias2018}, can be de-biased for the lower frequency of gaseous edge-on debris discs, yielding 0.17\% \citep[][Section 1.2]{2025MNRAS.537..229B}. Since most debris discs lack observable circumstellar gas and also involve later spectral types, the true rate is likely lower. Addressing the sample size and spectral type limitations, \citet{2025MNRAS.537..229B} explored the entire HARPS archive of 6100 stars, finding a 0.1\% detection rate overall, broadly consistent with the de-biased estimate. However, this rises to 0.4\% when considering only A-type stars, with both main detections ($\beta$\,Pictoris and HD\,172555) being young systems already known to host exocomets. The large age range among the observed A-type stars in HARPS suggests youth enhances detectability of Ca\,{\sc ii} absorption variability.}

Beyond pre-selection and sample biases, traditional spectroscopic detection rates suffer from a methodological flaw. \re{The detection rate is computed as the number of detections divided by the number of targets searched, but this approach does not take into account differences in observational cadence or the widely varying transit frequencies across systems.} While $\beta$\,Pictoris shows transits on any given night, HD~172555 does not, making the latter likely more representative of typical systems where longer monitoring increases detection probability. Properly accounting for the typical number of nights per target, \citet{2025MNRAS.537..229B} derive $1.4 \times 10^{-4}$ detections in Ca\,{\sc ii} per star per night, or 0.05 per star per year. Given the compounding observational biases (pre-selection of favourable targets, brighter stars, and inconsistent monitoring), current exocomet occurrence rates in spectroscopy remain highly uncertain.

\rev{In contrast, photometric surveys benefit from observing strategies that reduce target-selection bias.} For instance, \tess' all-sky strategy \rev{provides a magnitude-limited sample that includes many later-type stars, though it is still not volume-limited. The \kepler searches by \citet{Rappaport2018} and \citet{Kennedy2019} identified three systems with exocomet-like events among $\sim$150,000 stars observed for four years. This corresponds to roughly 600,000 star-years, yielding a detection rate of one event per 200,000 star-years (i.e. $\sim 7 \times 10^{-6}$ per star per year). \rev{The \tess-based search by \cite{2025MNRAS.542.1486N} finds a slightly higher rate of $\sim 3 \times 10^{-4}$ per star per year}, but both values are much lower than spectroscopic occurrence rates. At present, no definitive explanation has been proposed, although geometric factors may contribute: if photometric transits originate from exocomets at larger radial distances than those probed in spectroscopy, the decrease in transit probability with orbital distance could naturally produce lower detection rates \citep{Winn2010, Heller2024}.}%yields a magnitude-limited sample. While this sample is still biased compared to one that is volume-limited, it includes many later-type stars.
% completely eliminates any biases of spectral-type, stellar age, and stellar neighbourhood, and allowing us to re-assess the trends seen in spectroscopy. 

%The detections with \kepler in \citet{Rappaport2018} and \citet{Kennedy2019} yielded 3 systems with exocomet-like transits out of $\sim$150,000 stars observed, over four years. The cumulative observation time is therefore about 600,000 star-years, and the detection rate can be expressed as one detection per 200,000 stars per year of observation, or $5 \times 10^{-6}$ per star per year. The conclusions from a similar survey using \tess data find a \rev{slightly higher rate \citep{2025MNRAS.542.1486N}, but both rates are clearly much smaller than seen for spectroscopic observations, and at the time of writing both results are sufficiently recent that no explanations have been proposed}. If photometric transits are always for objects at greater distances than for spectroscopy, the transit probability could contribute to the difference \citep{Winn2010,Heller2024}.

% The work in TESS also resulted in new detections around giant stars, where 3 out of the 6 candidates in (Norazman in .prep) were identified around later-type stars. These are still small number statistics, however the recovery/detection of exocomet-like transits around younger stars is still evident with the detections in Kepler, $\beta$ Pic \citep{Zieba2019}, and the other half of detections in TESS (Norazman, et al. submitted), and hence the trend may indeed be real, and that the younger, more luminous stars are more likely to be exocomet host systems.

\subsection{Outlook for extant data}

\rev{A major opportunity for progress lies in exploiting the vast amount of existing photometric and spectroscopic data that has not yet been systematically searched for exocomets. While targeted studies and small-sample searches have identified the first examples, the volume of archival data now available far exceeds what has been analysed to date. Large, homogeneous searches, potentially assisted by machine-learning methods, are therefore the logical next step toward building an unbiased census of exocomet activity.}

\rev{Spectroscopic studies to date span many decades and a variety of instruments, but nearly all have been based on relatively small, biased, or pre-selected samples. Early searches focused on stars already known to show circumstellar variability \citep[e.g.][]{Lagrange1992,vidal-madjar1994,Lecavelier1997d} or debris-disc hosts \citep[e.g.][]{Montgomery2012,Welsh2018,Rebollido2020}. More recent work has begun mining larger archives (e.g. HARPS), but these datasets remain limited in wavelength coverage and dominated by later-type stars not optimally suited for detecting Ca\,{\sc ii} variability.}

\rev{A comprehensive, cross-instrument archival search, for example across HARPS, UVES, HIRES, ESPaDOnS, FEROS, and other high-resolution spectrographs, would vastly expand the explored parameter space. Such an effort would require consistent calibration across instruments with different resolutions, wavelength ranges, and SNR characteristics, but would open the door to the first truly unbiased assessment of time-variable absorption across a wide range of stellar types and ages. As with photometry, machine-learning tools may ultimately aid in building improved reference spectra, identifying subtle variable features, and quantifying detection significance in a reproducible way.}

\rev{Space-based missions such as \kepler, \ktwo, and \tess provide millions of high-precision light curves that contain far more information than has been extracted so far. Existing searches (e.g. those by \citealt{Rappaport2018}, \citealt{Kennedy2019}, and \citealt{2025MNRAS.542.1486N}) demonstrate that exocomet transits can be identified, but it remains unclear whether the approaches used so far recover are the most effective, and revisiting these datasets with alternative techniques may yield more detections.
Machine-learning approaches are particularly promising. Random Forest classifiers trained on statistical light-curve features \citep{Dobrycheva2023} and neural networks trained on synthetic transits \citep{Dumond2025} have already shown potential for improving sensitivity to non-periodic, asymmetric events. Applying and refining such methods to the full \kepler and \ktwo and ongoing \tess datasets could reveal a significantly larger population of exocomet candidates, while establishing more robust detection thresholds and false-positive controls.}

\section{Open Questions}\label{sec4}

Exocomet research faces numerous open questions, with the most fundamental currently being one of detection. Despite growing interest, the number of known exocometary systems remains relatively small. Many of these systems are considered candidates based on limited observations. For spectroscopic data, candidate systems are often derived from biased samples designed to maximise detection rates, and where biases are less, detection is significantly harder for later type stars (e.g. activity in Ca\,{\sc ii}, lower UV flux). While photometric detections lack such biases, they are exceedingly rare. Both spectroscopic and photometric studies indicate that young, early-type stars are overrepresented among exocomet candidate systems \citep[e.g.,][]{Strom2020, Iglesias2025}. This may align with expectations that younger and more luminous stars are more likely to host detectable exocomets. Biases complicate this picture, but it may nonetheless reflect a genuine trend \citep{Kennedy2019,2025MNRAS.537..229B}. These biases, combined with the scarcity of known exocomet systems, \rev{add uncertainty to the currently understood exocomet detection rates and ultimately affect in-depth study of the phenomenon.}
% present significant obstacles to statistical analysis and in-depth study of the phenomenon.

Beyond the challenge of detection, there is a broader ambition to achieve a deeper understanding of exocomets, akin to what has been achieved for Solar System comets. While such an understanding is ambitious—especially given the direct exploration of Solar System comets—the discovery of interstellar visitors like 1I/'Oumuamua, 2I/Borisov, \rev{and 3I/ATLAS} suggests a link between these objects and the bodies observed as transiting exocomets. Future missions to explore such visitors could provide valuable insights. For exocometary systems around other stars, $\beta$\,Pictoris stands out as an exemplary testbed. This system likely represents the best-case scenario for advancing in-situ exocomet science and serves as a foundation for developing techniques that might be applicable to other systems, if similar systems were found.

Below, we outline several key areas where progress is critically needed to advance our understanding of exocomets.

\subsection{How common are exocomets?}

\re{Understanding the frequency and distribution of exocomet activity across planetary systems is fundamental to constraining their dynamical and compositional evolution.} Their prevalence provides insight into the stability and architecture of planetesimal reservoirs, the efficiency of dynamical processes that perturb small bodies toward the inner system, and the overall frequency of active minor bodies in different stellar environments. As noted above, exocometary science is severely limited by a lack of systems with robust and repeated detections. Aside from the conclusion based on photometry that transiting exocomets are seen for of order 1:100,000 stars per year of observation \citep{Kennedy2019}, and the possibility that they are more prevalent for young early-type stars, little is known. However, to improve our understanding of their occurrence rates, greater precision and observations of larger numbers of stars are required, both in photometry and spectroscopy.

\subsubsection{Photometry
}

The upcoming \plato mission \rev{is one of the most anticipated missions this decade and} will play an important role \rev{for potential exocomet detections}. Scheduled for launch at the end of 2026, \plato will be a large-scale space-based photometric survey with substantial advantages over previous missions. \rev{Its key strengths are} high photometric precision and a wide field of view. For bright stars, \plato can achieve a precision of $\sim$7\,ppm (Tmag = 4) in two hours using its 26 cameras working together  \citep{Matuszewski2023,2024MNRAS.535.1778E}.  This \rev{design increases both reliability} and allows for cross-verification of signals: astrophysical phenomena will be detected by multiple cameras, while single-camera detections can be attributed to instrumental errors.

\rev{The other} main advantage of \plato is its observing time. \plato will provide two \rev{main} options for its observing strategy: the Long-Duration Observation Phase (LOP), which \rev{would observe targets for up} to three years, and the Step-and-Stare Phase (SOP) \rev{that observes targets} up to one year. Both strategies surpass the observing durations of \tess and are more comparable with \kepler. Furthermore, \plato’s focus on bright stars ($V_\mathrm{mag} \leq 8.5$ for its brightest targets, with its main sample between 8 and 11 magnitudes) \rev{should} enable detection rates higher than those of \tess, \cheops or \kepler \citep{2025ExA....59...26R}. By observing over 245,000 FGK-stars \citep{Matuszewski2023}, \plato is uniquely positioned to expand our understanding of the prevalence of exocomets and uncover how common these objects truly are. The high precision and long observation durations should yield high-quality data with new exocomet detections and valuable statistics.

\rev{The Roman Telescope is another instrument that may yield exocomet detections with a different perspective. Roman is set to launch in 2027 and is a wide-field, space-based survey, observing targets in the near-infrared (NIR). While Roman's NIR sensitivity will be less responsive to the smallest dust grains of an exocometary tail, it will be sensitive to the larger grains. This wavelength-dependent response means that exocomet transits observed by Roman would likely appear shallower than observations in the optical, but the combination of optical and NIR observations can provide powerful constraints on dust grain size distributions \citep{Kalman2024,Gibson2025}. Therefore, observations with Roman introduce multi-wavelength photometric coverage of exocomets, allowing for complementary science alongside \plato and \tess.}

\rev{Finally, Earth 2.0 \citep{Ge2022} is a mission concept currently planned for launch around 2027, and may also be an interesting avenue for exocomet detections. It aims to follow the legacy of \kepler, discovering Earth-like planets around Sun-like stars using high-precision space-based photometry. Earth 2.0 aims to continuously monitor a region that strongly overlaps with the original \kepler field, and is expected to complement the original \kepler planets as well as new detections. The Earth 2.0 cameras are also predicted to have even better photometric precision compared to \kepler and potentially \plato, and so any data products available could be particularly valuable for exocomet observations.}

% Multi-facility exocomet searches such as that carried out for 5 Vul with TESS and CHEOPS (Rebollido et al. 2023) provide another pathway to increase SNR and obtain color information. In particular, high-cadence TESS observations will be complemented in the South by high-precision but low-cadence color information from LSST with the VRO (K´alm´an et al. 2024). Furthermore, exocomet host stars are likely to overlap the large fields of view of the PLATO long-duration stares (K´alm´an et al. 2024) and Roman Galactic Bulge Time Domain Survey, providing high-cadence, high-precision multi-color and NIR observations, respectively -- text from Gibson 2025.

\subsubsection{Spectroscopy
}
Spectroscopic searches also hold significant potential for advancing exocomet research. As noted above a critical step forward would be a comprehensive search of archival optical spectra to quantify the exocomet detection rate, e.g. in Ca\,{\sc ii}. A more ambitious way forward would be a major effort at UV wavelengths where atomic species are more readily detectable. The primary challenge with UV observations lies in securing sufficient observation time on HST to observe a stellar sample of sufficient size and diversity to draw valuable conclusions. A systematic search could target early- and mid-spectral-type stars in nearby young moving groups, with multi-epoch observations providing valuable insights.
Such a search might, for example, aim for multi-epoch observations of early and mid-spectral type stars in each of the nearby young moving groups. Given the expense and pressure of HST, such a sample might be geometrically biased in favour of detection, but in a way that should not be correlated with exocomet existence, by favouring stars more likely to be edge-on rotators. 

\rev{Looking ahead, the upcoming Ariel mission \citep{Tinetti2021} may provide an opportunity for exocomet characterisation through multi-band observations. Primarily designed to study exoplanet atmospheres, Ariel provides a combination of synchronous visual and infrared photometry alongside low-resolution NIR spectroscopy (up to $7.8\,\mu m$), which could enable confirmation of dusty transits while constraining their chemical composition, providing important parameters for exocomet modelling and confirming the nature of these exocomet bodies.}

% ARIEL —The upcoming Ariel mission (Tinetti et al. 2021) will be observing individual exoplanet systems from a target list. Ariel is planned to have the ability of synchronous visual + infrared photometry (VISPhot, FGS1 and FGS2; Szab´o et al. 2022) together with NIR spectra taken by two low-resolution spectrographs (NIRSpec and AIRS Tinetti et al. 2021) with coverage up to 7.8 µm, which is also capable of synthetic spectrophotometry. Ariel will be the first dedicated exoplanet space observatory with multiband abilities, and will be very much suited for a confirmation of the comet nature of dusty transients. Moreover, the low-resolution spectra from Ariel can indeed reveal the chemical constituents of the dusty transients, providing an important constraint to the deeper modeling and confirming the origin of them. -- text from Kalman2024.

\subsection{What physical \re{dust} properties could we potentially observe?}

Here we will primarily focus on dust properties, which are currently completely unconstrained. This limitation arises because to date all space-based transit surveys are monochromatic, meaning that there is in fact no empirical evidence that the dimming seen during photometric transits is actually dust. Based on our understanding of the Solar system comets dust is a likely cause, and confirmation might be provided with simultaneous observations in two filters. Such an observation is however extremely challenging; obtaining 0.1\% level photometry is possible, but difficult, and would be made even more complicated by attempts to observe a bright $V=3.9$\,mag star. Nonetheless, attempting to constrain or measure the colour dependence of exocomet transits to learn about dust properties is a clear priority, with \plato's ``fast'' cameras being an obvious candidate instrument \citep{2025ExA....59...26R}.

\subsubsection{Polarimetry}

Polarimetry provides a powerful method to study the characteristics of scattered light by particles, offering insight into their microphysical properties. This technique is highly sensitive to the physical parameters of scatterers, meaning that even slight changes in these properties can significantly affect the observed degree of linear polarisation.

%Using the phase-angle dependence of the linear polarization degree, it is possible to infer the albedo, refractive index (real and imaginary parts), and details about the size, porosity, and structure of dust particles. For example, Umow's law \citep{Umow1905} allows albedo estimation from the maximum polarization value. Similarly, the relationship between albedo and the polarimetric slope, derived from the polarization maximum, is well-established \citep{Kenknight+1967, Widorn1967}.

\begin{figure}
    \centering
    \includegraphics[angle=270,width=\linewidth]{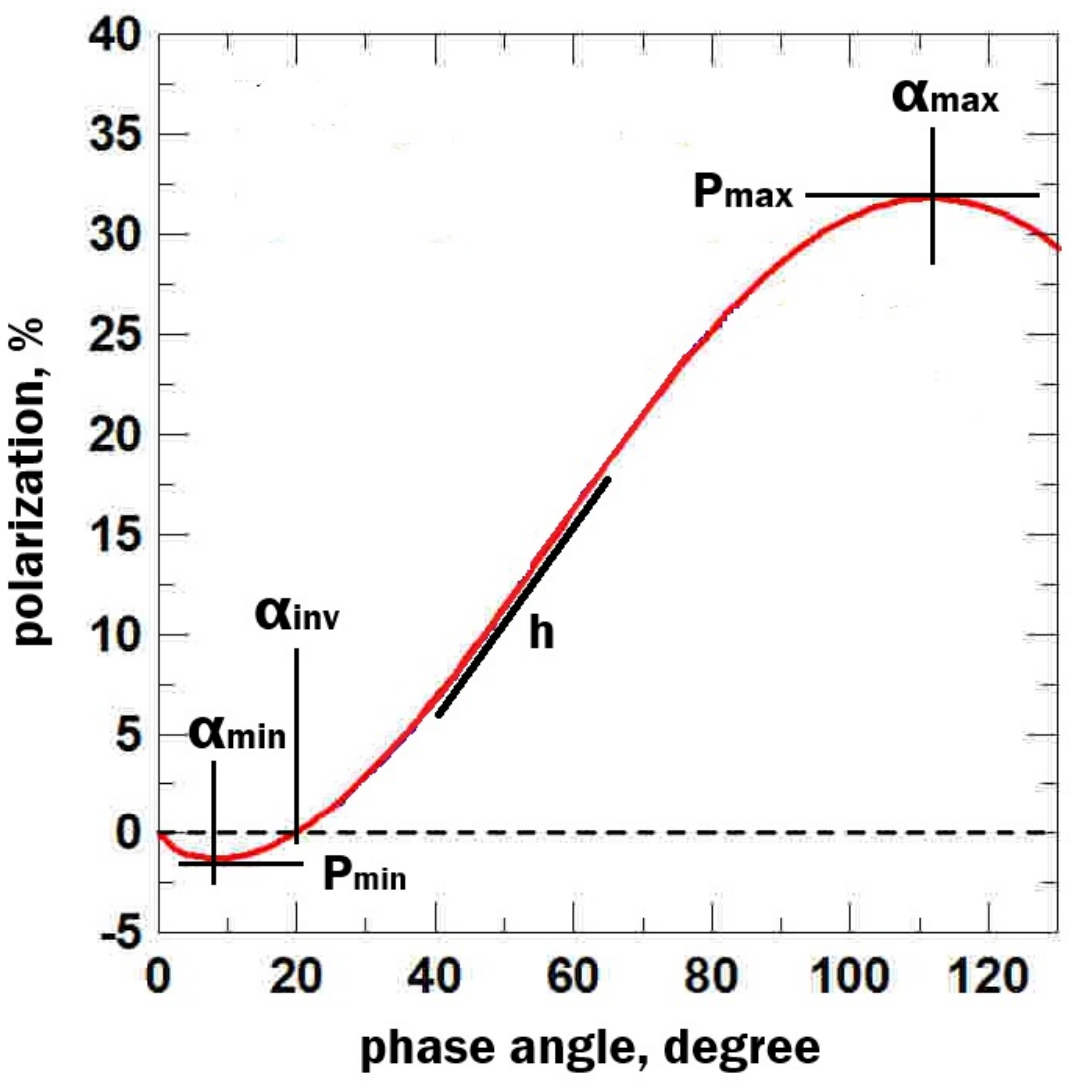}
    \caption{Synthetic phase-angle dependence of the linear polarisation degree in a form suggested by \cite{1993LPICo.810..194L}.}
    \label{fig:polarimetry}
\end{figure}

\rev{The phase-angle dependence of the linear polarisation degree has a specific wave-shaped form with so-called negative and positive branches (see Fig.~\ref{fig:polarimetry}). Its parameters can be used to estimate some characteristics of dust grains. For instance, the polarisation maximum ($P_\mathrm{max}$) is related to albedo \citep{Umow1905} and, likewise, to grain size \citep{1989A&A...213..469D}; the slope (h) is connected to albedo and indicates a very low albedo with a high own value \citep{1989A&A...213..469D}. The microstructure of grains is correlated with the inversion angle ($\alpha_\mathrm{inv}$); the fluffier structures produce the largest $\alpha_\mathrm{inv}$, while compact powders reduce $\alpha_\mathrm{inv}$ \citep{1989A&A...213..469D}. However, these relations are challenging because they require individual phase-angle dependence across the entire angle range, which is almost impossible for a single object. Therefore, one usually uses measurements for several objects. Furthermore, dust characteristics, including refractive index or chemical composition, could be reproduced using different models. The main idea of these methods is to simulate the scattering properties of dust, including linear polarisation degree, using materials with varying refractive indices, grain shapes, porosities, and sizes. In this case, the observed measurements are used for validating the modeled values. For more details, we refer readers to, for example, \citet{2015psps.book..379K,2004come.book..577K,2024AJ....168..164K,2024MNRAS.528.7027S,2011JQSRT.112.1848Z}, and references therein.}

In the Solar System, a complete phase-angle dependence is often required to obtain precise physical parameters from cometary polarisation measurements. However, achieving such a \rev{coverage} is challenging \rev{because most comets cannot be observed across the full $0^\circ$–$180^\circ$ phase-angle range. Similar geometric limitations are expected for exocomets. For transiting systems in particular, small phase angles, when the comet is behind the star, are entirely inaccessible, and the observable phase-angle range is restricted by the orbital inclination and the system viewing geometry.}

\rev{To date, polarimetry of exocomets remains a challenging task, especially for transiting events where the expected signal is extremely faint. Even for Solar System comets, a single polarimetric measurement is insufficient to derive robust physical properties, and repeated observations across multiple phase angles are generally required. For exocomets, individual transits will therefore not provide enough phase-angle coverage for full physical characterisation, but polarimetric measurements can still serve as a valuable additional diagnostic.}

\rev{Whether a transiting exocomet is observable in polarimetry depends strongly on the system configuration. In face-on debris-disc systems, detectable exocomets are most likely those with orbital inclinations close to the disc plane (near $0^\circ$ or $180^\circ$). In edge-on systems, exocomets with significantly inclined orbits are more favourable, as their scattered-light polarisation signal is more easily separated from the stellar contribution and from the bulk disc emission.}

\rev{In practice, the most promising current strategy is not time-resolved polarimetry of individual transits, but rather imaging polarimetry of the disc environment after an exocomet candidate has been identified. Instruments such as SPHERE/ZIMPOL on the VLT \citep{2019A&A...631A.155B} offer the necessary sensitivity and spatial resolution to reveal localised enhancements in polarised intensity. Such ``clumps'' in polarisation maps may correspond to dust released by an exocometary body, providing an opportunity to detect exocomet-related structures and study their scattering properties.}

\rev{In an ideal scenario, an exocomet would first be detected through photometric or spectroscopic variability, its orbit estimated, and targeted polarimetric observations obtained subsequently to search for excess polarised flux at the expected orbital location. Future high-contrast polarimeters on ELTs will greatly improve the achievable signal-to-noise ratio and may enable direct detection and characterisation of transient exocometary dust clouds.}

Despite these challenges, valuable estimates can still be made without a complete phase-angle dependence. Observations of Solar System comets are typically sparse, especially for long-period comets. By combining these sporadic data points with particle models, the observed polarisation degree can be reproduced, yielding estimates of physical parameters such as size, albedo, and porosity \citep{Dlugach+2018, Ivanova+2019, Ivanova+2021, Ivanova+2023}. For instance, Solar System comets are often modelled using mixtures of transparent particles (e.g., silicates) and absorbing particles (e.g., organics). Adjusting the ratio of these components enables accurate reproduction of polarisation degrees \citep[e.g.,][]{Kochergin+2021}. Polarimetric modelling has also been used to estimate dust particle characteristics, including refractive indices, based on their chemical composition \citep{Rosenbush+2020, 2024MNRAS.528.7027S}.

\subsubsection{Multi-wavelength studies}

The combination of polarimetric and multi-wavelength photometric observations can further refine these estimates. By incorporating dust colour information, the range of potential variables in simulations is constrained, allowing for more accurate determinations of physical parameters, such as the refractive index \citep{Shubina+2024aa}. Multi-wavelength data also help identify variations in dust composition and structure across different regions of the exocometary coma \citep{Lagrange1996,Lecavelier1999}. For example, colour indices derived from simultaneous photometric observations can reveal compositional gradients or particle size distributions, providing deeper insights into the physical and chemical processes shaping exocomets.

\rev{Dedicated observational programmes have attempted to detect exocomet signatures, or related circumplanetary or circumstellar dust structures, using multi-instrument campaigns. Most notably, the large $\beta$\,Pictoris Hill-sphere transit campaign \citep{Kenworthy2021,Zieba2024} combined bRing, ASTEP, BRITE, HST/STIS, \tess, and ground-based facilities in a concerted attempt to detect dust or gas associated with $\beta$\,Pictoris b. Although no circumplanetary material was found, the programme demonstrated the feasibility of coordinated, multi-instrument monitoring for detecting transient dusty or gaseous structures. Although no programme has yet yielded a confirmed simultaneous exocomet detection, the field is beginning to benefit from multi-wavelength, multi-instrument observational strategies.}

Simultaneous UV to mid-infrared observations have proven especially valuable for characterising the properties of transiting material. Observations of WD1145+017, for instance, demonstrated that transit depths remain consistent across a wide range of wavelengths, from UV to 4.5$\mu$m, suggesting a lack of small dust particles ($\lesssim~1.5~\mu$m) if the material is optically thin \citep{Xu2018a}. This finding highlights the importance of using broad wavelength coverage to detect subtle differences in particle size distributions or identify atypical dust populations.

For exocomets, accumulating multi-wavelength and multi-method observations enables a more robust characterisation of exocomets, revealing details about particle size, composition, and orbital behaviour that are inaccessible through single-method studies. Expanding these datasets will be critical for advancing our understanding of exocometary systems, uncovering their physical properties, and contextualising their role within planetary systems.

\subsection{What are the compositions of exocomets?}

Composition is a fundamental property of planetesimals, reflecting the conditions of their formation environment and their subsequent history. One key aspect of comets is that they are typically ``pristine'', in that most of their lifetime before becoming sublimating bodies on eccentric orbits has been spent at large orbital separations. Comets can therefore provide information about the environment within the protoplanetary disc where the planets formed. Similarly, the compositions of exocomets could provide comparable insights, although this requires understanding their formation locations, which might be inferred through dynamical studies.

%With the exception of $\beta$\,Pictoris, almost nothing is known about exocomet compositions. While multiple species have been detected for $\beta$\,Pictoris, and a few for 49~Ceti and 51~Oph \citep{2016ApJ...824..126M,2002ApJ...568..343R}, most systems are either detected only in Ca\,{\sc ii} or broadband photometry. The planetesimals that probably represent the source regions of exocomets, observed as debris disks, are thought to outgas CO \citep[e.g.][]{1995Natur.373..494Z,2011ApJ...740L...7M,2014Sci...343.1490D}, but no other cometary species have yet been detected (aside from the atomic products C and O). This lack of compositional information for exocomets contrasts starkly with metal-polluted white dwarfs, where in most cases the only information is compositional.

%Efforts to analyze archival UV spectra of $\beta$\,Pictoris’ exocomets have yielded promising insights. Recent studies \citep{2024A&A...684A.210V,2024A&A...691A...2V} reveal Solar-like abundances of Ni+/Fe+ and Cr+/Fe+ in these exocomets.  Extending such methods to other systems, like 49~Ceti or new candidate systems, could broaden our understanding of exocomet compositions beyond $\beta$\,Pictoris and place them in a larger context.

Polluted white dwarfs, those displaying elements heavier than Helium in the atmosphere, provide a powerful way to directly constrain the compositions of extrasolar minor bodies, including exocomets. These systems provide insights into the bulk compositions of accreted material. Details of this can be found in these review papers \citep{JuraYoung2014, Xu2024b}. Looking ahead, there are a few ongoing and planned multi-object spectroscopic surveys, such as SDSS-V, DESI, and 4MOST, that will likely detect many more polluted white dwarfs and reveal the full range of the compositions of extrasolar minor bodies.

High-resolution optical spectroscopy, combined with photometric monitoring, offers another layer of detail. For WD~1145+017, the circumstellar absorption lines were observed to weaken during transits, enabling constraints on the geometry of the circumstellar gas and the structure of the transiting material \citep{Xu2019a}. This approach demonstrates how combining spectroscopic and photometric data can yield a comprehensive picture of the dynamics and composition of exocometary systems.

Simultaneous spectroscopy and photometry could play a crucial role in resolving key questions about exocomet compositions. As mentioned above, combined observations might address whether the exocomets detected through spectroscopy and photometry originate from the same families or represent distinct populations. Additionally, they could clarify whether differences in detection rates arise from observing separate gas and dust components or reflect intrinsic compositional variations. Such studies will be vital for determining whether these techniques complement each other or reveal fundamentally different aspects of exocometary systems.

\section{Summary and Outlook}

Here we have outlined the history and status of exocomet detection, focussing on individual transiting bodies. While these are primarily detected around main-sequence stars, some are also seen in white dwarf systems. Both photometry and spectroscopy are used, with the latter detecting a variety of species (e.g. Ca, Mg, Al, Fe) at both optical and UV wavelengths. While it is expected that the photometric transits are due to dust in the tail/coma of the comet, there is not yet empirical evidence of reddening or polarisation that might be expected.

The techniques used are ultimately all based on time-based differencing, with spectroscopy comparing variation in specific absorption lines, and photometry looking for changes in the flux of a star over time. The major difference in these searches is that spectroscopy is so far based on individually targeted observations, while photometry uses large-scale surveys. Both methods yield a few to a few tens of detections, meaning that the derived occurrence rates are wildly different; an estimated rate for Ca\,{\sc ii} in spectroscopy is 0.05 per star per year, while for photometry it is $5 \times 10^{-6}$ per star per year. The reason for this difference is unclear, though both physics and transit probabilities may be at play.

Spectroscopic simulations may provide insights into future lines of inquiry for exocometary transit spectra. Simulating transit spectra for an exocomet as seen transiting, for instance, $\beta$\,Pictoris, could provide expected transition frequencies and relative transit depths for species beyond Ca and Na by assuming elemental abundances as measured in solar system comets \citep{Rubin2019}. Such efforts would require a new combination of tools, including fluorescence models of cometary ions \citep[e.g.,][]{Bromley2021}, a full treatment of ionisation and recombination processes (including photo- and collisional ionisation) to compute ion densities along the line of sight, and a radiative transfer engine such as the NASA Planetary Spectrum Generator \citep{2018JQSRT.217...86V,Villanueva2022} to compute the transit spectra. %We refer the reader to \cite{Vrignaud2025} for a detailed review of the processes governing ion populations in exocometary comae. A key challenge for future simulation efforts is determining the density and temperature profiles expected for exocomets, although recent work by \cite{2024A&A...684A.210V} has demonstrated an approach through measurement of Fe\,{\sc ii} column densities in exocomets around $\beta$\,Pictoris. 
Combining \rev{approaches as presented in \citet{2024A&A...684A.210V} with new modelling and higher-quality UV data will help identify previously unrecognised features in exocomet spectra and guide the design of future observational programmes.}%all of these tools in future studies may provide simulations which serve to spark searches for previously unidentified features in exocomet spectra, as well as inform proposals for newly designed observations.

In any field, it is likely that the most extreme objects are detected first, and that given sufficient interest and subsequent work that others follow. Exocomet science is so far different, in that $\beta$\,Pictoris' exocomets were first observed by \citet{Ferlet1987}, but no systems that are remotely comparable have yet been detected. This disconnect does not appear to be due to a lack of effort, as millions of stars have been searched in photometry and around 5,000 in spectroscopy. One of the major tasks for the future is therefore to find the next highly active exocomet system.

Addressing all current and future questions requires advancements in observational techniques and a deeper exploration of existing and new datasets. With upcoming missions and increasingly sophisticated observational methods, the field of exocomet research stands poised for significant breakthroughs, shedding light on the dynamic processes that shape planetary systems beyond our own.

Beyond what we have considered here, there are other future possibilities. A mission tasked with finding Earth-like planets will likely also have the capability to detect other bodies and/or material in the habitable zones of other stars. Indeed, exo-Zodiacal dust, which might be deposited by exocomets, has been a possible issue for such missions if the increased noise is too high \citep[e.g.][]{2012PASP..124..799R}. As suggested by \citet{2023A&A...671A.114J}, such a mission should also be able to image exocomets. \rev{Ultimately, exocomets represent a bridge between planetary formation, debris disc evolution, and the delivery of volatiles to emerging worlds. Continued progress will rely on combining photometric and spectroscopic searches, high precision modelling, and next-generation instruments capable of resolving faint and transient signals. By expanding the sample of known systems and improving our understanding of their physical and dynamical properties, exocomet research will continue to play a central role in revealing how planetary systems form, evolve, and sustain the small body populations that shape them.}

\backmatter

%\bmhead{Supplementary information}

%If your article has accompanying supplementary file/s please state so here. 

\bmhead{Acknowledgements}

We gratefully acknowledge support by the International Space Science Insitute, ISSI, Bern, for supporting and hosting the workshop on ``Exocomets: Bridging our Understanding of Minor Bodies in Solar and Exoplanetary Systems'', during which this work was initiated in July 2024.

The authors acknowledge support from the Swiss NCCR PlanetS and the Swiss National Science Foundation. This work has been carried out within the framework of the NCCR PlanetS supported by the Swiss National Science Foundation under grants 51NF40182901 and 51NF40205606. 

J.K. acknowledges support from the Swedish Research Council (Project Grant 2017-04945 and 2022-04043) and of the Swiss National Science Foundation under grant number TMSGI2\_211697.
RBW is supported by a Royal Society grant (RF-ERE-221025).
AN is supported by the University of Warwick and the Royal Society.
OSh is grateful for the support funded by the EU NextGenerationEU through the Recovery and Resilience Plan for Slovakia under the project No. 09I03-03-V01-00001.
D.I. acknowledges support from the Science and Technology Facilities Council via grant number ST/X001016/1.

\section*{Declarations}

\bmhead{Conflict of interest/Competing interests} Not applicable.

\bibliography{sn-bibliography}% common bib file

\end{document}